\documentclass[fleqn]{article}
\usepackage{amsmath,amssymb}
\usepackage{epsfig,graphicx}

\catcode`@=11 \@addtoreset{equation}{section} \catcode`@=12

\begin{document}
\title{\vskip-1.7cm \bf  New type of hill-top inflation}
\date{}
\author{A.O.Barvinsky$^{1,\,2,\,3}$, A.Yu.Kamenshchik$^{4,\,5}$ and D.V.Nesterov$^{1}$}
\maketitle
\hspace{-8mm} {\,\,$^{1}$\em Theory Department, Lebedev Physics Institute, Leninsky Prospect 53, Moscow 119991, Russia}\\
{$^{2}$\em Tomsk State University, Department of Physics, Lenin Ave. 36, Tomsk 634050, Russia}\\
{$^{3}$\em Pacific Institue for Theoretical Physics, Department of Physics and Astronomy, University of}\\
{\em British Columbia, 6224 Agricultural Road, Vancouver, BC V6T 1Z1, Canada}\\
{$^{4}$\em Dipartimento di Fisica and INFN, via Irnerio 46, 40126 Bologna, Italy}\\
{$^{5}$\em L.D.Landau Institute for Theoretical Physcis, Kosygin str. 2,
119334 Moscow, Russia}\\

\begin{abstract}
We suggest a new type of hill-top inflation originating from the initial conditions in the form of the microcanonical density matrix for the cosmological model with a large number of quantum fields conformally coupled to gravity. Initial conditions for inflation are set up by cosmological instantons describing underbarrier oscillations in the vicinity of the inflaton potential maximum. These periodic oscillations of the inflaton field and cosmological scale factor are obtained within the approximation of two coupled oscillators subject to the slow roll regime in the Euclidean time. This regime is characterized by rapid oscillations of the scale factor on the background of a slowly varying inflaton, which guarantees smallness of slow roll parameters $\epsilon$ and $\eta$ of the following inflation stage. A hill-like shape of the inflaton potential is shown to be generated by logarithmic loop corrections to the tree-level asymptotically shift-invariant potential in the non-minimal Higgs inflation model and $R^2$-gravity. The solution to the problem of hierarchy between the Planckian scale and the inflation scale is discussed within the concept of conformal higher spin fields, which also suggests the mechanism bringing the model below the gravitational cutoff and, thus, protecting it from large graviton loop corrections.
\end{abstract}

\section{Introduction}
Once very popular the problem of initial conditions in cosmology \cite{noboundary,tunnel} starts  attracting attention again. In particular, it raises the issue of a possible origin of inflation in the Starobinsky $R^2$-gravity \cite{Starobinskymodel} and the non-minimal Higgs inflation model \cite{BezShap} which are widely regarded as explaining the CMB data \cite{WMAP,Planck} and incorporating a possible link between these data and the phenomenology of the particle Standard model \cite{we,BezShap1,Wil,BezShap3,RGH}. Three known sources of cosmological initial conditions are pure no-boundary \cite{noboundary} and ``tunneling" \cite{tunnel} quantum states of the Universe and the Fokker-Planck equation for coarse-grained cosmological evolution \cite{Fokker-Planck}. Unfortunately, observer independent treatment of the no-boundary state leads to an insufficient amount of inflation which starts at the minimum of the inflaton potential rather than its maximum. Volume weighting \cite{volume-weighting} or top-down approach \cite{replica} to the no-boundary state seem to resolve this issue but remain with the problem of consistency of complex tunneling geometries and normalizability of the quantum ensemble in cosmology. Tunneling state has a rather uncertain ground based on the hyperbolic rather than Schroedinger nature of the Wheeler-DeWitt equation. Moreover, no-boundary and tunneling states do not have a clear canonical quantization ground and within the Euclidean path integral construction represent a special quasi-vacuum state. Fokker-Planck equation leads to eternal inflation, Multiverse concept, Boltzmann brains, etc. and raises the issue of a very contrived measure \cite{sad}.

The model that circumvents the above problems and is based on first principles of canonical quantization was suggested in \cite{slih,why}. It represents the synthesis of two main ideas -- a new concept of the cosmological microcanonical density matrix as the initial state of the Universe and application of this concept to the system with a large number of quantum fields conformally coupled to gravity. It plays important role within the cosmological constant and dark energy problems. In particular, its statistical ensemble is bounded to a finite range of the effective cosmological constant \cite{slih}, it incorporates inflationary stage and is potentially capable of generating the cosmological acceleration phenomenon within the so-called Big Boost scenario \cite{bigboost}. Moreover, the conformal cosmology provides perhaps the first example of the initial quantum state of the inflationary Universe, which has a thermal nature of the primordial power spectrum of cosmological perturbations. This suggests a new mechanism for the red tilt of the CMB anisotropy \cite{DGP/CFT,CMBA-theorem}, complementary to the conventional mechanism which is based on a small deviation of the inflationary expansion from the exact de Sitter evolution \cite{ChibisovMukhanov}.

This setup has a clear origin in terms of operator quantization of gravity theory in the Lorentzian signature spacetime and is based on a natural definition of the microcanonical density matrix as a projector on the space of solutions of the quantum gravitational Dirac constraints -- the system of the Wheeler-DeWitt equations \cite{why,whyBFV}. Its statistical sum has a representation of the Euclidean quantum gravity path integral \cite{slih,why}
    \begin{eqnarray}
    &&Z=
    \!\!\int\limits_{\,\,\rm periodic}
    \!\!\!\! D[\,g_{\mu\nu},\varPhi\,]\;
    e^{-S[\,g_{\mu\nu},\varPhi\,]},         \label{Z}
    \end{eqnarray}
over metric $g_{\mu\nu}$ and matter fields $\varPhi$ which are
periodic on the Euclidean spacetime with a time compactified to a circle $S^1$.

This statistical sum has a good predictive power in the Einstein theory with the primordial cosmological constant $\varLambda$ and the matter sector which mainly consists of a large number $\mathbb{N}$ of quantum fields $\phi$ conformally coupled to gravity \cite{slih,why} with the action $S_{CFT}[\,g_{\mu\nu},\varPhi\,]$. The dominant contribution of numerous conformal modes allows one to overstep the limits of the usual semiclassical expansion. Integrating these modes out
one obtains the effective gravitational action with $S_{CFT}[\,g_{\mu\nu},\varPhi\,]$ replaced by the quantum effective action of the conformal fields $\varGamma_{CFT}[\,g_{\mu\nu}]$. On the Friedmann-Robertson-Walker (FRW) background, $g_{\mu\nu}dx^\mu dx^\nu=N^2(\tau)\,d\tau^2
    +a^2(\tau)\,d^2\Omega^{(3)}$,
with a periodic lapse function $N(\tau)$ and scale factor $a(\tau)$ -- the functions of the Euclidean time belonging to the circle $S^1$ \cite{slih},
this action is exactly calculable by using the local conformal transformation to the static Einstein universe and the well-known trace anomaly,
    \begin{eqnarray}
    &&g_{\mu\nu}\frac{\delta
    \varGamma_{CFT}}{\delta g_{\mu\nu}} =
    \frac{1}{4(4\pi)^2}g^{1/2}
    \left(\alpha \Box R +
    \beta E +
    \gamma C_{\mu\nu\alpha\beta}^2\right),     \label{anomaly}
    \end{eqnarray}
which is a linear combination of Gauss-Bonnet $E=R_{\mu\nu\alpha\gamma}^2-4R_{\mu\nu}^2+ R^2$, Weyl tensor squared $C_{\mu\nu\alpha\beta}^2$ and $\Box R$ curvature invariants with spin dependent coefficients. The resulting $\varGamma_{CFT}[\,g_{\mu\nu}]$ turns out to be the sum of the anomaly contribution and the contribution of the static Einstein universe -- the Casimir and free energy of conformal matter fields on the sphere $S^3$ at the temperature determined by the circumference of the compactified time dimension $S^1$. This is the main calculational advantage provided by the local Weyl invariance of $\varPhi$ conformally coupled to $g_{\mu\nu}$.

Solutions of equations of motion for the full effective action, which give a dominant contribution to the statistical sum, are the cosmological instantons of $S^1\times S^3$ topology. These instantons serve as initial conditions for the cosmological evolution $a_L(t)$ in the physical Lorentzian spacetime. The latter follows from $a(\tau)$ by analytic continuation $a_L(t)=a(\tau_*+it)$ at the point of the maximum value of the Euclidean scale factor $a_+=a(\tau_*)$. The fact that these instantons exist only in the finite range of $\varLambda$ implies the restriction of the microcanonical ensemble of universes to this range, which is akin to the solution of the string landscape problem if one assumes that this model is a low-energy limit of the string theory.

As was mentioned in \cite{bigboost} this scenario can incorporate a finite inflationary stage when $\varLambda$ is replaced by a composite operator $\varLambda(\phi)=V(\phi)/M_P^2$ -- the potential of the conformally non-invariant scalar field $\phi$ which is slowly varying during the Euclidean and inflationary stages and decaying in the end of inflation by a usual exit scenario. The goal of this paper is to develop such a generalization to the model with the dynamical inflaton $\phi$,
    \begin{eqnarray}
    &&S[\,g_{\mu\nu},\phi,\varPhi\,]=
    \int d^4x\,g^{1/2}\,\left(-\frac{M_P^2}2\,R
    +\frac12(\nabla\phi)^2+V(\phi)\right)
    +S_{CFT}[\,g_{\mu\nu},\varPhi\,],         \label{tree1}
    \end{eqnarray}
whose potential $V(\phi)$ simulates the effect of the primordial cosmological constant. With this generalization the restriction of the cosmological term range, $\varLambda(\phi)=V(\phi)/M_P^2$, becomes a selection of the range of $\phi$ or fixation of the initial conditions for inflation. These initial conditions is a principal goal of this paper.

As we will see, this conformal cosmology realizes the initial conditions in the form of the new type of hill-top inflation originating from the underbarrier oscillations of the inflaton $\phi$ and the scale factor $a$ in the vicinity of local {\em maxima} of $V(\phi)$. Thus, these initial conditions provide a considerable amount of inflation with the parameters calculable from the saddle point configurations for the microcanonical partition function. The qualitative picture of the hill-top inflation is shown on Fig.1 -- the inflaton slowly rolling from the potential hill in real Lorentzian time originates from its underbarrier oscillations after the Euclidean-Lorentzian transition at the turning point $\phi_*$. This is, however, not a tunneling in the usual sense, because there is no classically allowed state before this tunneling, and the cosmological instanton in the underbarrier regime is just the saddle point of the microcanonical partition function.

\begin{figure}[h]
\centerline{\epsfxsize 11cm \epsfbox{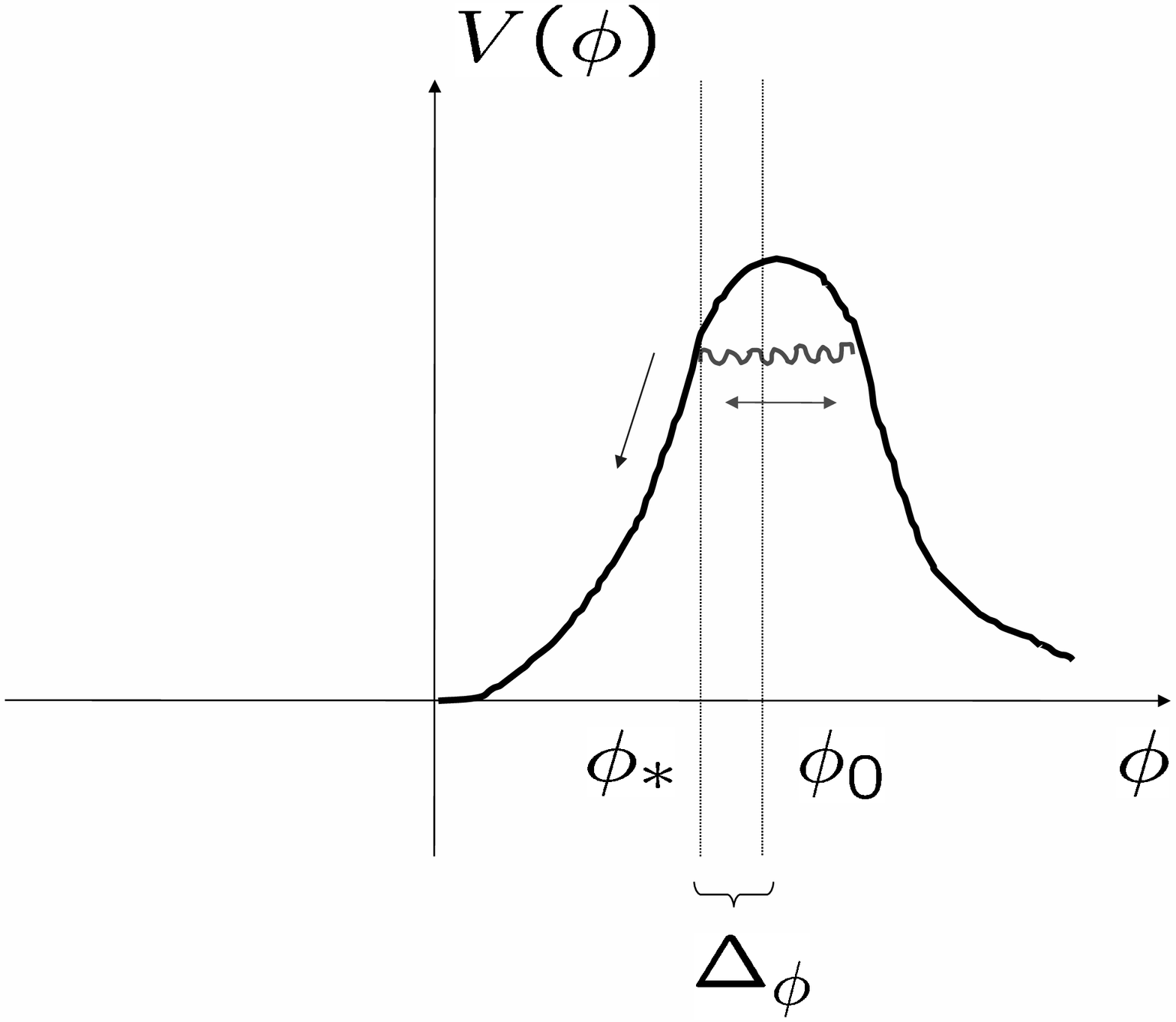}}
\caption{\small Picture of hill-top inflation: underbarrier oscillations indicated by waverly line give rise to inflationary slow roll at the turning point $\phi_*$.
 \label{Fig.1}}
\end{figure}

\section{The model with the fundamental cosmological constant}

For cosmology with $S^1\times S^3$ topology the statistical sum (\ref{Z}) is dominated by FRW solutions of the effective equation which in the gauge $N=1$ reads \cite{slih,why}
    \begin{eqnarray}
    &&-\frac{\dot a^2}{a^2}+\frac{1}{a^2}
    -B \left(\,\frac{\dot a^4}{2a^4}
    -\frac{\dot a^2}{a^4}\right) =
    \frac\varLambda3+\frac{C}{ a^4},\quad
    \dot a=\frac{da}{d\tau},               \label{efeq}\\
    &&C =
    \frac{B}2+\frac1{6\pi^2M_P^2}\,
    \frac{dF}{d\eta}, \quad B=\frac{\beta}{8\pi^2M_P^2}.  \label{bootstrap}
    \end{eqnarray}
This is the modification of the Euclidean Friedmann equation by the trace anomaly $B$-term and the radiation term $C/a^4$, both generated by $\varGamma_{CFT}[\,g_{\mu\nu}]$. The constant $B$ is defined by the coefficient $\beta$ of the Gauss-Bonnet term in (\ref{anomaly}) and $C$ here characterizes the sum of the {\em renormalized} Casimir energy $B/2$ and the thermal energy of CFT particles whose free energy $F(\eta)$ is a boson or fermion sum over field modes with energies $\omega$ on a unit 3-sphere,
    \begin{eqnarray}
    &&F(\eta)=\pm\sum_{\omega}\ln\big(1\mp
    e^{-\omega\eta}\big),\quad
    \eta=\int_{S^1} \frac{d\tau N}a,     \label{period}
    \end{eqnarray}
$\eta$ playing the role of the inverse temperature --- a period of the $S^1\times S^3$ instanton in units of the conformal time (\ref{period}). Eq.(\ref{efeq}) is derived by varying with respect to $N$ the effective action which was obtained by integrating the conformal anomaly (\ref{anomaly}) and independently taking into account the contribution of the auxiliary static Einstein universe.\footnote{\label{footnote1}This equation is independent of the anomaly coefficients $\alpha$ and $\gamma$, because it is assumed that $\alpha$ is renormalized  to zero by a local counterterm, $\varGamma_{CFT}\to\varGamma_{CFT}^R\equiv\varGamma_{CFT}+(\alpha/384\pi^2)\int d^4x\,g^{1/2}R^2$. This guarantees the absence of higher derivative terms in (\ref{efeq}) -- non-dynamical (non-ghost) nature of the scale factor -- and simultaneously gives the renormalized Casimir energy a partiular value proportional to $B/2=\beta/16\pi^2M_P^2$ \cite{universality}. Both of these properties are critically important for the instanton solutions of (\ref{efeq}). The coefficient $\gamma$ of the Weyl tensor term $C^2_{\mu\nu\alpha\beta}$ does not enter (\ref{efeq}) because $C_{\mu\nu\alpha\beta}$ identically vanishes for any FRW metric.}

The solutions of this integro-differential equation form the set of periodic $S^3\times S^1$ instantons with the oscillating scale factor -- {\em garlands} that can be regarded as the thermal version of the Hartle-Hawking instantons \cite{slih,why,DGP/CFT}. The scale factor oscillates $m$ times ($m=1,2,3,...$) between the maximum and minimum values (\ref{apm}), $a_-\leq a(\tau)\leq a_+$,
    \begin{eqnarray}
    a_\pm^2\equiv\frac{1\pm\Delta}{2H^2},
    \quad H=\sqrt{\frac\varLambda3},\quad \Delta=\sqrt{1-4CH^2}, \label{apm}
    \end{eqnarray}
so that the full period of the conformal time (\ref{period}) is the $2m$-multiple of the integral between the two neighboring turning points of $a(\tau)$, $\dot a(\tau_\pm)=0$, $\eta=2m\int_{a_-}^{a_+} da/\dot a a$. Thus, modulo the discrete number $m$, the instanton period is not arbitrary, but rather uniquely determined by the turning points of the scale factor evolution. In its turn, this conformal time period $\eta$ determines the effective temperature $T=1/\eta$ as a function of $M_P^2$ and $\varLambda$. This is the artifact of a microcanonical ensemble in cosmology \cite{why} with only two freely specifiable dimensional parameters --- the gravitational and cosmological constants.\footnote{As discussed in \cite{why}, the total energy of a {\em closed} cosmology as any other global conserved charge is identically zero, so that it cannot serve as an argument of the microcanonical statistical sum unlike in theories with nonzero conserved Hamiltonians and freely specifiable total energy.}

\begin{figure}[h]
\centerline{\epsfxsize 14cm \epsfbox{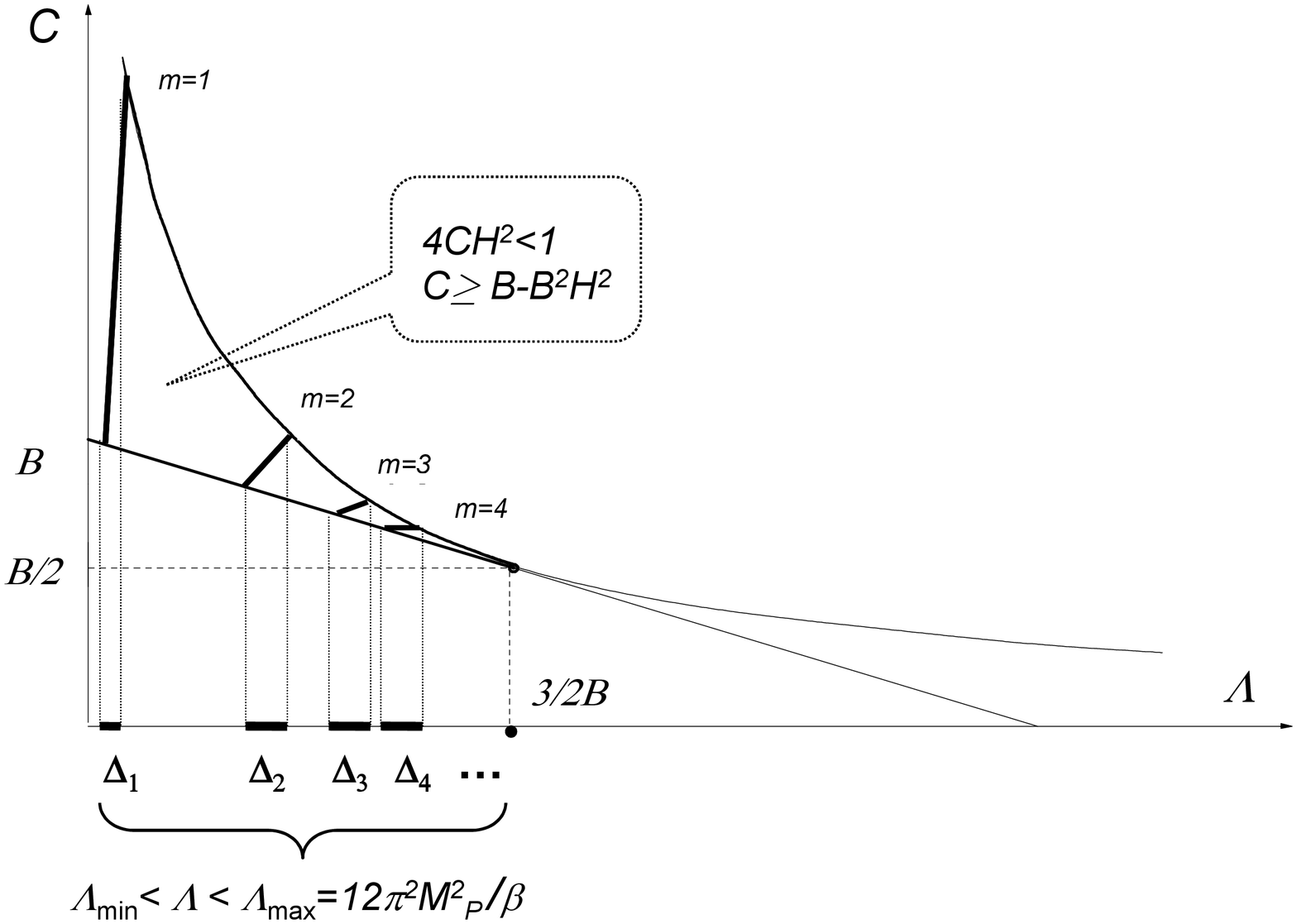}}
\caption{\small Band range of garland instantons
 \label{Fig.2}}
\end{figure}

These garland-type instantons exist only in the limited range of the cosmological constant $\varLambda=3H^2$ \cite{slih}.
As shown in \cite{slih}, the requirement of a periodic motion with turning points at $a_\pm$ restricts the set of these instantons to the curvilinear domain in the two-dimensional plane of $\varLambda$ and the amount of radiation constant $C$ (each instanton being represented by a point in this plane),
    \begin{eqnarray}
    C\geq B-B^2H^2, \quad C\leq \frac1{4H^2}, \label{triangle}
        \end{eqnarray}
which in terms of parameters $\Delta$ and $\varepsilon=1-2BH^2$ reads as $\Delta\leq\varepsilon\leq 1$. In this domain they form a countable, $m=1,2,...$, sequence of one-parameter families -- curves interpolating between the lower straight line boundary $C=B-B^2H^2$ and the upper hyperbolic boundary $C=1/4H^2$. Each curve corresponds to a respective $m$-fold instantons of the above type. Therefore, the spectrum of admissible values of $\varLambda$ has a band structure, each band $\Delta_m$ being a projection of the $m$-curve to the $\varLambda$ axis. The infinite sequence of bands of ever narrowing widths with $m\to\infty$ accumulates at the upper bound of this range $H^2_{\rm max}=1/2B$ or
    \begin{eqnarray}
    \varLambda_{\rm max}=\frac{12\pi^2 M_P^2}\beta. \label{Lambdamax}
        \end{eqnarray}
The lower bound $H^2_{\rm min}$ -- the lowest point of $m=1$ family -- can be obtained numerically for any field content of the model.

For a large number of conformal fields $\mathbb{N}$, and therefore a large $\beta\propto \mathbb{N}$, the both bounds are of the order $M_P^2/\mathbb{N}$. Thus the restriction (2.5) suggests that by specifying a sufficiently high number of conformal fields one can achieve a primordial value of $\varLambda$ well below the Planck scale where the effective semiclassical theory applies, but high enough to generate a sufficiently long inflationary stage.\footnote{This conclusion might signify that the energy scale of the model is too close to the widely accepted semiclassical cutoff $\sim M_P^2/\mathbb{N}$ in gravitational models with multiple species \cite{cutoff} to guarantee validity of perturbation theory. The resolution of this controversy is briefly discussed in Conclusions within the concept of conformal higher spin fields which are likely to resolve the hierarchy problem in conformal cosmology.}

\section{Dynamical inflaton in conformal cosmology}

Generalization to the model with the composite cosmological term (\ref{tree1}) implies the replacement of $\varLambda$ in the effective Friedmann equation (\ref{efeq}) by the function of $\phi$ and $\dot\phi$, $\varLambda\to\big(V(\phi)-\dot\phi^2/2\big)/M_P^2$, and the addition of the dynamical equation for $\phi$,
    \begin{eqnarray}
    \frac{d}{d\tau}\,
    \big(\,a^3\dot\phi\,\big)
    -a^3\frac{\partial V}{\partial\phi}=0.    \label{infleq}
    \end{eqnarray}

The possibility of periodic solutions of this system follows from counting its integration constants. Three integration constants $\phi_0$, $\dot\phi_0$, $a_0$ and the unknown period $T$ should satisfy three equations of periodicity for $\phi$, $\dot\phi$, $a$. In addition, at the nucleation point $\tau_*$ of the Lorentzian solution from the Euclidean instanton both time derivatives $\dot\phi_*$ and $\dot a_*$ should vanish, which gives one more equation, and this guarantees the existence of needed solutions at least up to some global obstructions. This means that, if some approximate periodic solution satisfying the condition of Euclidean-Lorentzian junction exists, then perturbation theory will guarantee the periodicity of perturbed solution.

Periodicity imposes an important restriction on the shape of the potential $V(\phi)$ -- integration of (\ref{infleq}) over the full period gives
    \begin{eqnarray}
    \oint d\tau\,a^3\frac{\partial V}{\partial\phi}=0.
    \end{eqnarray}
This implies that the gradient of $V$ changes its sign at some point $\phi_0$, $\partial V(\phi_0)/\partial\phi_0=0$, so that this is either a local minimum or maximum of the potential. In the vicinity of a minimum of $V(\phi)$ inflaton oscillations between the two turning points can only be in the over-barrier regime, that is in the real {\em Lorentzian} time. But from generic properties of the solution for the scale factor we know that it is periodic in the real {\em Euclidean} time. This means that inflaton oscillations should take place also in the real Euclidean time, that is in the under-barrier regime between the two turning points lying to the left and to the right of the potential {\em maximum}. Therefore, local minima of $V(\phi)$ are ruled out, and these oscillations are confined to the vicinity of the inflaton potential maximum (maxima).

This is a major conclusion that distinguishes density matrix initial conditions in microcanoical state conformal cosmology from those of the no-boundary wavefunction prescription. The latter, within the observer-independent treatment, favor minima of the inflaton potential and thus undermine inflationary predictions, because small initial values of $V(\phi)$ cannot provide us with a sufficient amount of inflation. Volume weighting and anthropic argumentation are then needed to handle the situation \cite{volume-weighting,replica} which, however, leave the ``objective" observer-independent distributions unnormalizable and hardly consistent within semiclassical expansion.

The effective Hubble factor $H^2=(V(\phi)-\dot\phi^2/2)/3M_P^2$ is no longer a constant parameter, and its time dependence is caused in view of (\ref{infleq}) by the friction term,
    \begin{eqnarray}
    &&\frac{d}{d\tau}H^2=
    3\frac{\dot a}a\,\dot\phi^2,
    \end{eqnarray}
which is small for slowly varying inflaton and scale factor and can be treated by perturbation theory. The leading order of this perturbation theory is given by the approximation of two coupled oscillators considered in much detail in \cite{slihinfl}. For the potential expanded near its maximum at $\phi_0$,
    \begin{eqnarray}
    V(\phi)\simeq V(\phi_0)-\frac12\,
    \mu^2 (\phi-\phi_0)^2,\quad \mu^2=
    -\frac{d^2V(\phi_0)}{d\phi_0^2}
    \equiv-V''_0>0,                  \label{quadratic}
    \end{eqnarray}
nearly harmonic oscillations of $\phi$ with a small amplitude $\Delta_\phi=\phi_0-\phi_*$, where $\phi_*$ is a turning point shown on Fig.1, are accompanied by oscillations of $a^2$
    \begin{eqnarray}
    &&\phi=\phi_0-\Delta_\phi\cos(\mu\tau),           \label{phi}\\
    &&a^2=\frac{1+\Delta\cos(\omega\tau)}{2H^2},\quad
    \omega=\frac{2H}{\sqrt{1-2BH^2}}\,.                   \label{omega}
    \end{eqnarray}
Such a behavior of the scale factor holds for the solutions close to the upper boundary of the domain (\ref{triangle}), when the scale factor is pinched between nearly coincident $a_\pm$ defined by (\ref{apm}), $\Delta\ll 1$. This follows from Eq.(\ref{efeq}) which in the parametrization  $a^2=(1+y\Delta)/2H^2$ reduces to the harmonic oscillator equation for the variable $y$, $\dot y^2+\omega^2y^2=\omega^2$, provided the following bound holds \cite{slihinfl}
    \begin{eqnarray}
    \Delta\ll\varepsilon\equiv 1-2BH^2.           \label{approximation0}
    \end{eqnarray}
For this to hold the parameters $\varepsilon$, $\Delta$ and $H$ -- functions of slowly oscillating $\phi$ -- must have a small change during the oscillation period $2\pi/\omega$, $\dot H\ll\omega H$, $|\dot\Delta|\ll\omega\Delta$. Together with smallness of the friction term in the inflaton oscillator, $|3\dot a\dot\phi/a|\ll |\ddot\phi|$, this imposes the bounds $\mu\gg\omega\Delta$ and
    \begin{eqnarray}
    W\equiv\frac{\dot\phi^2}{2M_P^2H^2}
    \sim \frac{\Delta_\phi^2}{M_P^2}
    \frac{\mu^2}{H^2}\ll\Delta.                      \label{W}
    \end{eqnarray}

Interestingly, the last bound -- smallness of $\dot\Delta$ -- can be replaced by the opposite limit $W\gg\Delta$ when Eq.(\ref{efeq}) leads to a strongly anharmonic $y$, $y\simeq\left|\,\sin\frac{\omega\tau}2\right|$, but still having the same period $2\pi/\omega$, provided a large function $W$ is slowly varying, $\dot W\ll\omega W$ \cite{slihinfl}. This leads to an extra bound on $\mu$, $\mu\ll\omega$, so that the frequency of inflaton oscillations $\mu$ belongs to a limited but nonempty range
    \begin{eqnarray}
    \omega\gg \mu\gg \omega\Delta.       \label{bound4}
    \end{eqnarray}

The motion of two coupled oscillators with frequencies $\mu$ and $\omega$ is periodic when $m\mu=n\omega$ with integer $m$ and $n$, and $m\gg n$ in the above range of $\mu$ -- fast oscillations of $a$ with a slower motion of $\phi$. Apart from the requirement of a valid (an)harmonic oscillator approximation, this range of $m\gg n$ can be justified by the slow roll regime for the inflationary stage. Slow roll parameters at the Euclidean-Lorentzian transition point $\eta_*\equiv M_P^2V''_*/V_*=-\mu^2/3H^2$ and $\epsilon_*\equiv (M_P V'_*/V_*)^2/2=\frac12(\Delta_\phi/M_P)^2 (\mu^2/3H^2)^2$ are related by the equation
    \begin{eqnarray}
    \epsilon_*=
    \frac12\left(\frac{\Delta_\phi}{M_P}\right)^2
    \eta_*^2,                                      \label{epsvseta}
    \end{eqnarray}
which implies for $\Delta_\phi\sim M_P$ a typical relation $\epsilon_*\sim\eta_*^2\ll\eta_*$ characteristic of the Starobinsky model \cite{Starobinskymodel} or the model with a non-minimally coupled inflaton \cite{we,nonminimal}. Now, in view of (\ref{omega}) it follows that
    \begin{eqnarray}
    H^2=\frac1{2B+4n^2/\mu^2m^2},    \label{Hmn}
    \end{eqnarray}
and
    \begin{eqnarray}
    &&|\,\eta_*|=\frac23\,B\mu^2+\frac43\frac{n^2}{m^2},  \label{etamn}
    \end{eqnarray}
which is not small unless $B\mu^2\ll 1$ and $m^2\gg n^2$.

For $m\gg n\geq 1$ the number $m$ of the $a$-oscillations during the full period implies the $m$-fold garland instanton with the Hubble parameter close to the upper bound of its range \cite{slih}
    \begin{eqnarray}
    H^2\simeq\frac1{2B}
    \left(1-\frac{\ln^2 m^2}{2\pi^2m^2}\right). \label{Hcrit}
    \end{eqnarray}
The Euclidean solutions in this limit can be called slow-roll ones, because the rate of change of the scale factor is much higher than that of the inflaton field. The Euclidean version of the slow-roll regime is, however, rather peculiar, because in contrast to the Lorentzian case with monotonically changing variables here the scale factor and inflaton are oscillating functions of time. The details of these solutions can be found in \cite{slihinfl}. Here we give a simplified overview of these solutions and their relation to the conventional slow roll parameters of inflation in the Lorentzian theory.

Comparison of (\ref{Hcrit}) with (\ref{Hmn}) implies that $n\simeq \sqrt{B\mu^2}\,\ln m/\pi$, so that in view of $n\geq 1$ the lower bound on $m$ is exponentially high, and the ratio $n/m$ in (\ref{etamn}) is exponentially small, $n/m\leq \exp(-\pi/\sqrt{B\mu^2})$. As a result the solution becomes very close to the upper quantum gravity scale -- the cusp of the curvilinear triangle on Fig.1 and the corresponding slow roll smallness parameter (\ref{etamn}) expresses in terms of the quantity $B\mu^2=B|\,V_*''|$,
    \begin{eqnarray}
    H^2\simeq\frac1{2B},\quad
    \eta_*\simeq-\frac23\,B\mu^2.    \label{parameters}
    \end{eqnarray}
In view of the known CMB data for $n_s=1-6\epsilon_*+2\eta_*\simeq 1+2\eta_*\simeq 0.96$, this quantity is also very small, $B\mu^2\sim 0.01$. Now, with
     $\varepsilon\simeq 2(\ln m/\pi m)^2
    \leq 2\exp(-2\pi/\sqrt{B\mu^2})/B\mu^2$,
the requirement of the harmonic oscillator approximation (\ref{approximation0}), $\Delta\ll\varepsilon$, establishes the range of the amplitude of scale factor oscillations $\Delta$. In terms of the slow roll smallness parameter $\eta_*\sim-0.01$ it reads
    \begin{eqnarray}
    \Delta\ll
    \frac1{|\eta_*|}
    e^{-2\sqrt2\pi/\sqrt{3|\eta_*|}}. \label{Deltabound1}
    \end{eqnarray}
Below we will estimate the bound on the first smallness parameter $\epsilon_*$ in conformal cosmology from the requirement that it should model initial conditions for  non-minimal Higgs inflation. This requires the knowledge of the amplitude of the inflaton field oscillations $\Delta_\phi$. As we will see,  this amplitude will have a sub-Planckian value, so that in view of (\ref{epsvseta}) a typical relation $\epsilon_*\sim\eta_*^2$ characteristic of the Starobinsky model or the model of non-minimal Higgs inflation will hold and signify that $\epsilon_*$ adds a negligible contribution to the CMB spectral parameter and provides a very small tensor to scalar ratio $r=16\epsilon_*$.

\section{The shape of the potential: non-minimal Higgs inflation and $R^2$ model}
Microcanonical state conformal cosmology can serve as a source of initial conditions for the non-minimally coupled Higgs inflation model \cite{BezShap,nonminimal,Gorbunov} which, together with the Starobinsky model of $R^2$-inflation \cite{Starobinskymodel}, is considered as one of the most promising models fitting the CMB data \cite{WMAP,Planck}. There is a twofold reason for that because, firstly, the Higgs inflation model has a natural mechanism of forming a hill-like effective potential and, secondly, it provides a relation $\varepsilon_*\sim\eta_*^2\ll|\eta_*|$, $\eta_*<0$, which establishes a strong link between the observable value of the CMB spectral parameter $n_s=1-6\epsilon_*+2\eta_*\simeq 1+2\eta_*$ and the value of the Higgs mass discovered at LHC \cite{we,BezShap1,Wil,BezShap3}. This relation, as we will see below, will be provided by the bound on the amplitude of the inflaton oscillations $\Delta_\phi$ in the underbarrier regime.

The inflationary model with a non-minimally coupled Higgs-inflaton $H$, $\varphi^2=H^\dag H$, as any other semiclassical model, has a  low-derivative part of its effective action (appropriate for the inflationary slow-roll scenario),
    \begin{equation}
    \varGamma_{\rm Higgs}[\,g_{\mu\nu},\varphi\,]=\int d^{4}x\,g^{1/2}
    \left(V(\varphi)-U(\varphi)\,R(g_{\mu\nu})+
    \frac12\,G(\varphi)\,
    (\nabla\varphi)^2\right)\ .   \label{effaction}
    \end{equation}
Its coefficient functions contain together with their tree-level part the logarithmic loop corrections with the UV normalization scale $\mu$,
    \begin{eqnarray}
    &&V(\varphi)=\frac\lambda{4}\,\varphi^4+
    O\Big(\,\varphi^4\ln\frac{\varphi^2}{\mu^2}\,\Big),\quad
    U(\varphi)=\frac{M_P^2+\xi\varphi^{2}}2+
    O\Big(\,\varphi^2
    \ln\frac{\varphi^2}{\mu^2}\,\Big),           \label{effPlanck}\\
    &&G(\varphi)=1+O\Big(\,
    \ln\frac{\varphi^2}{\mu^2}\,\Big).           \label{phirenorm}
    \end{eqnarray}
The coefficients of logarithms are determined by quantum loops of all particles and represent beta functions of the corresponding running coupling constants -- quartic self-coupling $\lambda$, non-minimal coupling of the Higgs field to curvature $\xi$ -- and anomalous dimension of $\varphi$.

Inflation and its CMB are easy to analyze in the Einstein frame of fields $\hat g_{\mu\nu}$, $\phi$, in terms of which the action (\ref{effaction}) $\varGamma_{\rm Higgs}[\,g_{\mu\nu},\varphi\,]=\hat\varGamma_{\rm Higgs}[\,\hat g_{\mu\nu},\phi\,]$ has a minimal coupling of the inflaton to curvature, $\hat U=M_P^2/2$, a canonical normalization of the inflaton field, $\hat G=1$, and a new inflaton potential, $\hat{V}(\phi)=M_P^4 V(\varphi)/4U^2(\varphi)$. Due to leading logarithmic terms this potential starts decreasing to zero for large values of $\varphi\simeq M_P\,\exp(\phi/\sqrt{6}M_P)/\sqrt\xi$, $\hat{V}(\phi)\sim
1/\ln(\varphi^2/\mu^2)\to 0$, $\varphi\to\infty$. Of course, this behavior cannot be extrapolated to infinity, because semiclassical expansion fails at transplanckian  energies, but the maximum of the potential, which is reached at some $\bar\phi$, $\hat{V}'(\bar\phi)=0$, corresponds for large $\xi\sim 10^4$ (matching CMB data with the Higgs mass value \cite{we,RGH}) to a subplanckian scale $\hat{V}(\bar\phi)\sim M_P^4/96\xi^2\ll M_P^4$.
Therefore the potential starts bending down in the semiclassical domain and acquires a shape of the hill suitable for our hill-top inflation scenario.

In fact a similar qualitative behavior holds in any order of loop expansion, because in the leading logarithm approximation of $l$-th loop order both $V(\varphi)$ and $U(\varphi)$ grow like $l$-th power of the logarithm while their ratio $V/U^2$ in the Einstein frame potential decreases like $(\ln(\varphi^2/\mu^2))^{-l}$. This property was confirmed numerically within RG resummation of leading logarithms in \cite{RGH} in the model of non-minimal Higgs inflation.

Application of our hill-top scenario to non-minimal Higgs inflation implies a transition between the original Jordan frame and the Einstein frame on the FRW background. We start with the full action containing the Einstein-Hilbert part, non-minimal Standard model part and large $\mathbb{N}$ sector of conformal fields
    \begin{eqnarray}
    S[\,g_{\mu\nu},H,...\varPhi\,]=
    S_{EH+SM}[\,g_{\mu\nu},H,...\,]+
    S_{CFT}[\,g_{\mu\nu},\varPhi\,],
    \end{eqnarray}
where
    \begin{eqnarray}
    S_{EH+SM}[\,g_{\mu\nu},H,...\,]=
    \int d^{4}x\,g^{1/2}
    \left(\frac{\lambda\varphi^4}4
    -\frac{M_P^2+\xi\varphi^2}2\,R+
    \frac12\,(\nabla\varphi)^2+...\right)
    \end{eqnarray}
includes the contributions of the Higgs field $\varphi^2\equiv H^\dag H$ non-minimally coupled to the metric and of the other Standard model fields denoted by ellipses. Quantization of this theory in the original {\em Jordan} frame\footnote{Quantization of scale invariant theories in the Einstein frame with the asymptotically shift invariant inflaton potential, allegedly, stabilizes radiative corrections \cite{BezShapSib}. Flatness of the potential, however, results not only in the smallness of vertices, but also makes the theory effectively massless. This undermines the gradient expansion which is a corner stone of the Higgs inflation model and brings to life strong infrared effects \cite{IR}. This and the other properties of shift-invariant potentials in the Einstein frame make us to prefer quantization in the Jordan frame.}, results in the effective action of the gravitating Higgs model (\ref{effaction}) and the effective action of the conformal sector, $S_{EH+SM}[\,g_{\mu\nu},H,...\,]\to \varGamma_{\rm Higgs}[\,g_{\mu\nu},\varphi\,]$, $S_{CFT}[\,g_{\mu\nu}, \varPhi\,]\to\varGamma_{CFT}[\,g_{\mu\nu}]$.

If one rewrites the Higgs effective action in the Einstein frame of fields $\hat g_{\mu\nu}$ and $\phi$ on the FRW background, $\hat\varGamma_{\rm Higgs}[\,\hat g_{\mu\nu},\phi\,]|_{FRW}=\hat\varGamma_{\rm Higgs}[\,\hat a,\hat N,\phi\,]$, then it reads as a classical action but in terms of the Einstein frame scale factor $\hat a$ and lapse $\hat N$ and with quantum corrected potential $\hat V(\phi)$ of the above hill-like shape. Conversion of the effective action of conformal fields to the Einstein frame variables, $\varGamma_{\rm CFT}[\,a,N\,]=\hat\varGamma_{\rm CFT}[\,\hat a,\hat N,\phi\,]$, generates an additional contribution which can be interpreted as the effect of the conformal anomaly due to the transition from the Jordan frame to the Einstein frame. This contribution is, however, small for a slowly varying scalar field, $|\dot\phi|/M_P\ll|\dot{\hat a}|/\hat a$,
so that $\hat\varGamma_{\rm CFT}[\,\hat a,\hat N,\phi\,]\simeq
\varGamma_{\rm CFT}[\,\hat a,\hat N\,]$, and the full action $\hat\varGamma_{\rm Higgs}[\,\hat a,\hat N,\phi\,]+\varGamma_{\rm CFT}[\,\hat a,\hat N\,]$ takes the form of the minimally coupled inflaton action in terms of hatted variables and with the quantum corrected hill-top potential. One can apply to it the above analysis. The additional smallness restriction $|\dot a|/a\gg|\dot\phi|/M_P$ (we omit hats from now on) implies the bound
    \begin{eqnarray}
    \frac{\Delta_\phi^2}{M_P^2}\,
    \frac{\mu^2}{H^2}\ll
    \frac{\Delta^2}\varepsilon,  \label{cond3}
    \end{eqnarray}
which is stronger than (\ref{W}). On account of the first bound (\ref{bound4}), $\mu\gg\omega\Delta$ or $\Delta^2/\varepsilon\ll\mu^2/H^2$, it takes the form $\Delta_\phi^2/M_P^2\ll 1$ and, thus, decreases $\epsilon_*$ in (\ref{epsvseta}) even below its value in the Starobinsky or Higgsflation model. This makes the estimate for the spectral parameter $n_s$ even less sensitive to the value of $\epsilon_*$ and even further reduces the CMB tensor to scalar ratio $r=16\varepsilon_*$.

The above picture also applies to the case when the Higgs model is replaced with the Starobinsky model of the $R^2$-gravity, $S_{EH+SM}[\,g_{\mu\nu},H,...\,]\,\to\, S_\xi^{\rm Star}[\,g_{\mu\nu}]=\frac12\int d^{4}x\,g^{1/2}(-M_P^2\,R-\xi\,R^2/2)$, because it can be reformulated as Einstein gravity with a non-minimally coupled scalar field
    \begin{equation}
    S^{\rm Star}_\xi[\,g_{\mu\nu},\varphi\,]=\int d^{4}x\,g^{1/2}
    \left\{-\frac{M_P^2}2\,
    \left(1+\xi\frac{\varphi^2}{M_P^2}\right)R
    +\frac{\xi\varphi^4}4\,\right\}.
    \end{equation}
Transition to the Einstein frame,  $1+\xi\varphi^2/M_P^2=
\exp(2|\,\phi\,|/M_P\sqrt6)$, $g_{\mu\nu}=\exp(-2|\,\phi\,|/M_P\sqrt6)\,
\hat g_{\mu\nu}$, recovers the action (\ref{tree1}) of the minimally coupled and canonically normalized inflaton $\phi$ with the potential $\hat V=M_P^4
(1-e^{-2|\phi|/M_P\sqrt6})^2/4\xi$.
If the slow roll condition (\ref{cond3}) holds, then this transition as above leaves the action of the model unchanged. Therefore, it takes the form of the original effective action with the minimal dynamical inflaton having this particular potential. It is asymptotically shift invariant at large $\phi$, but logarithmic radiative corrections render it a hill-top shape by the mechanism discussed above, and one can apply the hill-top scenario of the above type.

In fact, the inclusion of the Starobinsky model becomes indispensable if we put forward as a guiding principle a necessity to preserve garland instantons of the SLIH scenario \cite{slih}. This is because the $R^2$ term is the only means to render a non-dynamical -- and therefore stable non-ghost -- nature of the scale factor mode  and a particular value of the Casimir energy $\sim B/2$ in (\ref{efeq}) (see footnote {\ref{footnote1}). This can be expressed as a simple relation
\begin{equation}
S_\xi^{\rm Star}+\varGamma_{CFT}= S_{\xi+\alpha/96\pi^2}^{\rm Star}+\varGamma_{CFT}^R,
\end{equation}
where $\varGamma_{CFT}^R$ is the effective action of conformal fields with the trace anomaly (\ref{anomaly}) coefficient $\alpha$ renormalized to zero by adding the $R^2$ counterterm in contrast to the original action $\varGamma_{CFT}$ of the theory with a nonzero $\alpha$. Then the last equation can be interpreted as the way $R^2$-gravity plays its
double role -- part of it performs finite renormalization of $\alpha$ to zero, $\varGamma_{CFT}\to\varGamma_{CFT}^R$, while the rest of it generates a dynamical inflaton feeding the CFT scenario with the potential
    \begin{equation}
    \hat V=M_P^4\frac{\big(1-e^{-2|\phi|/M_P\sqrt6}\,\big)^2}
    {4\xi+\frac\alpha{24\pi^2}},                      \label{hatV}
    \end{equation}
which is converted by logarithmic quantum corrections to a hill like shape.

\section{Conclusions}
Thus, for a wide range of parameters satisfying the bounds (\ref{approximation0})-(\ref{bound4}) the microcanonical state cosmology with a dynamical inflaton possessing a hill-like potential has instantons which are very close to the solutions of the model with a primordial cosmological constant. These instantons are described by the approximation of two coupled oscillators and generate a new type of hill-top inflation depicted on Fig.1. Hill-top inflation histories in Lorentzian signature time, $\phi_L(t)$ and $a_L(t)$, originate by analytic continuation of the Euclidean solutions (\ref{phi})-(\ref{omega}) to $\tau=2m\pi/\omega+it$, where $m\gg 1$ is the number of oscillations of the scale factor in the garland instanton. For small time $t$ the linearized Lorentzian solutions read
    \begin{eqnarray}
    &&\phi_L(t)=\phi_0-\Delta_\phi\cosh(\mu t),\quad a^2_L(t)=\frac1{2H^2}+\frac{\Delta}{2H^2}\cosh(\omega t),\quad
    \omega t\ll 1,       \label{aL}
    \end{eqnarray}
while at later times nonlinear effects start dominating, so that the Lorentzian version of the nonlinear equation (\ref{efeq}) with a dynamical $\varLambda$, $\varLambda\to\big(V(\phi_L)+\dot\phi_L^2/2\big)/M_P^2 \equiv\rho_\phi/M_P^2$, enters the stage. This equation can be rewritten in the manifestly Friedmann form with the effective Planck mass $M_{\rm eff}(\rho)$ depending on the full matter density $\rho$ which together with $\rho_\phi$ includes the primordial radiation of the conformal cosmology \cite{bigboost},
    \begin{eqnarray}
    &&\frac{\dot a_L^2}{a_L^2}+\frac1{a_L^2}=
    \frac{\rho}{3M^2_{\rm eff}(\rho)}, \quad
    M^2_{\rm eff}(\rho)=\frac{M_P^2}2\left(\,1+\sqrt{1
    -\frac{\beta\,\rho}{12\pi^2M_P^4}}\,\right), \\
    &&\rho=\rho_\phi+\frac{R}{a^4_L}, \quad R=3M_P^2\Big(\,C-\frac{B}2\,\Big)=\frac1{2\pi^2}
    \sum\limits_\omega\frac\omega{e^{\eta\omega}\mp1}.
    \end{eqnarray}
The further evolution for large $t$ consists in the fast quasi-exponential expansion during which the primordial radiation gets diluted, the inflaton field decays by a conventional exit scenario and goes over into the quanta of conformally non-invariant fields produced from the vacuum.\footnote{A realistic model should contain a sector of non-conformal fields which can be negligible on top of conformal fields in the early Universe but eventually starts dominating in the course of cosmological expansion.} They get thermalized and reheated to give a new post-inflationary radiation having a sub-Planckian energy density, $\rho\to\rho_{\rm rad}\ll M_P^4/\beta$, so that $M_{\rm eff}\to M_P$, and one obtains a standard general relativistic inflationary scenario for which initial conditions were prepared by our garland instanton.

Exponentially high number of the instanton folds $m$, corresponding to the upper bound on the effective cosmological term $H^2= 1/2B$ in the range (\ref{triangle}), guarantees the smallness of slow roll parameters $\epsilon$ and $\eta$ beginning with their values $\epsilon_*$ and $\eta_*$ at the Euclidean-Lorentzian transition point. They turn out to satisfy the relations $\epsilon\sim\eta^2$ and $\eta<0$ characteristic of the well-known non-minimal Higgs inflation or Starobinsky $R^2$ gravity models. Hill-like potentials in these models can be generated by logarithmic loop corrections to their tree-level asymptotically shift-invariant potential, so that with the bound (\ref{cond3}) on the amplitude and frequency of oscillations of cosmological instantons, conformal cosmology can be regarded as a source of initial conditions for the Higgs inflation with a strong non-minimal coupling or Starobinsky $R^2$ gravity. The bound (\ref{cond3}) implies the relation $\epsilon\ll\eta^2$ at the onset of inflation, which even stronger bounds the tensor to scalar ratio in the CMB signal of these models.

A major difficulty with this scenario is the hierarchy problem. The inflation scale $H^2\sim1/2B=4\pi^2 M_P^2/\beta$ requires the parameter $\beta$ to be exceedingly large in order to match with the CMB data for the energy density $V_*=12\pi^2 M^4_P/\beta$ which for the non-minimal Higgs inflation should be about $10^{-11}M_P^4$ \cite{we}. The needed $\beta\sim 10^{13}$, which is not accessible in models with conformal fields of low spins $s=0,1/2,1$, might be due to the hidden sector of weakly interacting conformal higher spin (CHS) fields \cite{Vasiliev,GKPST}. Their partial contributions $\beta_s$ to $\beta=\sum_s\beta_s$ grow with spin as $s^6$ \cite{GKPST,Tseytlin}, so that the needed value can be attained with the tower of CHS fields, $0\leq s\leq S$, up to $S=100$ containing $\mathbb{N}\sim 10^6$ polarizations \cite{CHScosmology} or with the individual CHS field of the spin $s\simeq 200$ and $\mathbb{N}\sim 4\times 40000$. Quite interestingly, this coincides with the estimate for the average value of $\beta$ per one conformal degree of freedom, $\beta/\mathbb{N}\sim 10^6\div 10^9$,  at which the thermal correction to the CMB spectral parameter, $\Delta n_s^{\rm thermal}\sim -0.001$, appears in the third decimal order \cite{CMBA-theorem}.\footnote{This can be achieved by essentially reducing the number of garland folds $m$, which is in its turn possible for potentials whose convexity $V''_*$ at the Euclidean-Lorentzian transition is much smaller than $V''_0$.} This means that a potential resolution of the hierarchy problem via CHS simultaneously makes measurable the thermal contribution to the CMB red tilt, which is complementary to the conventional tilt of the CMB spectrum \cite{ChibisovMukhanov}.

Interestingly, CHS fields provide a new mechanism of stabilizing radiative corrections. As in any other effective theory, we do not have control over quantum effects in high-energy limit, and all our conclusions remain valid only below a certain cutoff. Since {\em graviton loops} in nonrenormalizable quantum gravity are suppressed by inverse powers of Planck mass $1/M_P^2$, this cutoff a priori coincides with the Planck scale. However, in view of the universality of the gravitational interaction for a large number of quantum species $\mathbb{N}$ this suppression factor becomes $\mathbb{N}/M_P^2$, and the cutoff scale reduces to $M_P/\sqrt{\mathbb{N}}$. This is a well known expression achieved on the basis of perturbative arguments \cite{Veneziano} or implications of the Hawking radiation from a semiclassical black hole \cite{cutoff}.  With this cutoff our predictions in conformal cosmology become questionable. Indeed, according to (\ref{Lambdamax}) the energy scale of the model $M_P/\sqrt\beta$ turns out to be too close to the cutoff and even can essentially exceed it provided $\beta\sim\mathbb{N}$.

Critical point with CHS fields, which resolves this difficulty, is that their $\beta$ grows nonlinearly with their total number of polarizations $\mathbb{N}$. According to \cite{GKPST,Tseytlin} partial $\beta_s$ and partial numbers of physical degrees of freedom of CHS particles $\mathbb{N}_s$ grow with the spin respectively as $s^6$ and $s^2$ (overall $\beta$ and $N$ for a CHS tower of a height $S\gg 1$, $0\leq s\leq S$, are correpondingly $S^7$ and $S^3$), so that the ratio of the cutoff scale to the actual energy scale of the conformal cosmology is $\sqrt{\beta/\mathbb{N}}\sim 10^3$. Thus, the model is in the quantum state three or four orders of magnitude below its gravitational cutoff, and the contribution of nonrenormalizable graviton loops is negligible.

It is important that the above derivation of the cutoff is based namely on the number of species $\mathbb{N}$, rather than the central charge $c\sim\beta$ -- one of the coefficients of the trace anomaly -- unlike in the approach of \cite{Veneziano} requiring the smallness of the {\em total} one-loop contribution relative to the tree-level part. In that approach the cutoff turns out to be $M_P/\sqrt\beta$ and is usually identified with $M_P/\sqrt\mathbb{N}$ in simple models with $\beta$ and $\mathbb{N}$ of the same order of magnitude. Here the cutoff $M_P/\sqrt\mathbb{N}$ follows from the smallness of only the graviton loop contribution, but not of the quantum species one. Therefore, the usually ignored difference between $\mathbb{N}$ and $\beta$ becomes critically important in the case of CHS fields with $\beta\gg\mathbb{N}$ and makes this mechanism possible \cite{CHScosmology}.

It remains to explain why do we expect stabilization of radiative corrections from the conformal sector. Its contribution is big, because it is weighted by $\beta\gg\mathbb{N}$, and this contribution is critically important because its dynamical balance against the tree-level part establishes the upper bound (\ref{Lambdamax}) on the energy scale of inflation and leads to a particular structure of the instanton solutions. But this sector is perturbatively renormalizable, and for linear fields on FRW background its contribution is exactly calculable via the one-loop conformal anomaly.\footnote{For simplicity we consider conformal fields without self-interaction, but the generalization to nonlinear case is straightforward. Due to renormalizability gravitational multi-loop counterterms are exhausted by the same curvature invariants as in the one-loop order, which generate the three-parameter trace anomaly (\ref{anomaly}), the main effect of nonlinearity being a slow logarithmic RG running of $\alpha$, $\beta$ and $\gamma$ \cite{Komargodski}.} Moreover, for the FRW background this sector is free from {\em logarithmic} UV divergences, because in Weyl invariant theories on this background the counterterms -- $\int d^4x\,g^{1/2}C_{\mu\nu\alpha\beta}^2$ and the Euler number of the $S^1\times S^3$ instantons (contributed by the Gauss-Bonnet invariant $E$) -- are both vanishing. Thus, this sector of the model is free from UV renormalization ambiguity, and its power divergences are absorbed by the renormalization of the cosmological and gravitational coupling constants. Note that the $R^2$-term of the Starobinsky model, considered above and used for a {\em finite} renormalization of the Casimir energy and simulation of the effective cosmological term, belongs to the UV {\em finite} sector of the Weyl invariant fields coupled to gravity. Therefore, elimination of higher-derivative ghosts by this finite renormalization cannot be broken by radiative corrections, and this satisfies the criterion of ``naturalness" in quantum field theory.

All this seems to justify a special role of CHS fields for the stability of the model against uncontrollable graviton loop effects. Though this mechanism of a large $\beta$ is vulnerable to criticism regarding fine tuning \cite{IR} and the problem of perturbative unitarity\footnote{Hidden CHS sector is coupled to the observable fields only gravitationally, but below the cutoff this coupling is suppressed similarly to the contribution of graviton loops, and this essentially reduces the effects of non-unitarity. A fundamental solution of the unitarity problem for higher-derivative CHS fields is anticipated only at the nonperturbative level under a better understanding of the gauge theory of interacting higher spins.} for CHS fields, it fits presently very popular ideas of string theory, higher spin gauge theory \cite{Vasiliev} and holographic duality \cite{GKPST}, and it will be considered in much detail in \cite{CHScosmology}.

We accomplish the paper with a brief discussion of another property of our model. This is an unnaturally high number of garland folds, $m\geq \exp(\pi\sqrt2/\sqrt{3|\eta|})$ dictated by the phenomenological smallness of the second slow roll parameter $\eta$ in (\ref{parameters}). This problem, however, can be solved by considering potentials more realistic than (\ref{quadratic}) when their convexity $|V''_*|$ at the transition point is much smaller than that at the their maximum $\mu^2=|V''_0|$, cf. footnote 6 above. Exponential sensitivity of the results at the high $m$ tail of the instanton range (\ref{triangle}) would then relax the bounds on $m$ and on $\Delta$. The final caveat regarding the role of inflationary smallness parameters in the CMB of our model is that its precise spectrum has not yet been found, except the thermal contribution to $n_s$ found in \cite{CMBA-theorem}. Conventional dependence of CMB parameters on $\epsilon$ and $\eta$ might be modified by a nontrivial sound speed and the effect of nonlocal quantum corrections in the effective equations for cosmological perturbations -- the issue subject to current research \cite{CFTCMB}.

\section*{Acknowledgments}
The authors are grateful to F.   Bezrukov, D. Gorbunov, V.Mukhanov, A. Panin, M. Shaposhnikov and S.  Sibiryakov for fruitful and thought-provoking correspondence and discussions. They also greatly benefitted from helpful discussions with J.B. Hartle, D. Blas, C. Burgess, C. Deffayet, P. Stamp, A.A. Starobinsky, N.Tsamis, A. Tseytlin, W. Unruh, A. Vilenkin and R. Woodard. A.B. is grateful for hospitality of Theory Division, CERN, and Pacific Institute of Theoretical Physics, UBC, where this work was initiated. This work was also supported by the RFBR grant No. 14-02-01173 and by the Tomsk State University Competitiveness Improvement Program.


\begin{thebibliography}{99}
\bibitem{noboundary}J. B. Hartle and S. W. Hawking, {\it  	
Wave Function of the Universe, Phys. Rev.}  {\bf D 28} (1983) 2960;\\
    S. W. Hawking, {\it  	
The Quantum State of the Universe, Nucl. Phys.} {\bf B 239} (1984)  257.

\bibitem{tunnel}A. D. Linde, {\it Quantum creation of an inflationary universe, J. Exp. Theor. Phys.}  {\bf 60} (1984) 211;\\
    V. A. Rubakov, {Particle Creation In A Tunneling Universe, J. Exp. Theor. Phys.  Lett.} {\bf 39} (1984) 107;\\
    Ya. B. Zeldovich and A. A. Starobinsky, {\it  	
Quantum creation of a universe in a nontrivial topology, Sov. Astron. Lett.} {\bf 10} (1984) 135;\\
    A. Vilenkin, {\it  	
Quantum Creation of Universes, Phys. Rev. } {\bf D 30} (1984) 509, {\it The Wave function discord, Phys. Rev.}  {\bf  D 58} (1998) 067301  [gr-qc/9804051].

\bibitem{Starobinskymodel} A. A. Starobinsky, {\it A New Type of Isotropic Cosmological Models Without Singularity, Phys. Lett. B} {\bf 91} (1980) 99.

\bibitem{BezShap}F. Bezrukov and M. Shaposhnikov, {\it The Standard Model Higgs boson as the inflaton, Phys. Lett.}  {\bf B 659}
(2008) 703 [arXiv:0710.3755].

\bibitem{WMAP}G.Hinshaw et al, {\it Nine-Year Wilkinson Microwave
    Anisotropy Probe (WMAP) Observations: Cosmological Parameter Results, Astrophys. J. Suppl.}  {\bf 180} (2009) 225 [arXiv:1212.5226];\\
    E. Komatsu et al., {\it  	
Five-Year Wilkinson Microwave Anisotropy Probe (WMAP) Observations: Cosmological Interpretation, Astrophys. J. Suppl.} {\bf 180} (2009) 330 [arXiv:0803.0547].

\bibitem{Planck}P. Ade et al. (Planck Collaboration), {\it Planck 2013 results. XVI. Cosmological parameters, Astron. Astrophys.} {\bf 571} (2014) A16 [arXiv:1303.5076]; {\it Planck 2013 results. XXII. Constraints on inflation, Astron. Astrophys.} {\bf 571} (2014) A22 [arXiv:1303.5082].

\bibitem{we}A. O. Barvinsky, A. Yu. Kamenshchik and A. A. Starobinsky,
{\it Inflation scenario via the Standard Model Higgs boson and LHC, JCAP} {\bf 11} (2008) 021  [arXiv:0809.2104].

\bibitem{BezShap1}F. L. Bezrukov, A. Magnin and M. Shaposhnikov, {\it  	
Standard Model Higgs boson mass from inflation, Phys. Lett.}  {\bf B 675} (2009) 88 [ arXiv:0812.4950].

\bibitem{Wil} A. De Simone, M. P. Hertzberg and F. Wilczek, {\it  	
Running Inflation in the Standard Model, Phys. Lett.} {\bf B 678} (2009) 1 [arXiv:0812.4946].

\bibitem{BezShap3}F. Bezrukov and M. Shaposhnikov, {\it  	
Standard Model Higgs boson mass from inflation: Two loop analysis, JHEP} {\bf 07} (2009) 089 [arXiv:0904.1537].

\bibitem{RGH}A. O. Barvinsky, A. Yu. Kamenshchik, C. Kiefer, A.A. Starobinsky and C. Steinwachs, {\it Asymptotic freedom in inflationary cosmology with a non-minimally coupled Higgs field, JCAP} {\bf 12} (2009) 003 [arXiv:0904.1698];  {\it Higgs boson, renormalization group, and naturalness in cosmology}, Eur. Phys. J. {\bf C 72} (2012) 2219 [arXiv:0910.1041].

\bibitem{Fokker-Planck}A. A. Starobinsky, {\it Stochastic De Sitter (inflationary) Stage In The Early Universe, Lect. Notes Phys.} {\bf 246} (1986) 107;\\
    A. A. Starobinsky and J. Yokoyama, {\it Equilibrium state of a selfinteracting scalar field in the De Sitter background, Phys. Rev.} {\bf D 50} (1994) 6357 [astro-ph/9407016];\\
    C.P. Burgess, R. Holman, G. Tasinato and M. Williams, {\it EFT Beyond the Horizon: Stochastic Inflation and How Primordial Quantum Fluctuations Go Classical, JHEP} {\bf  03} (2015) 090 [arXiv:1408.5002];\\
    V. Vennin and A. A. Starobinsky, {\it Correlation Functions in Stochastic Inflation, Eur. Phys. J.}  {\bf C 75}  (2015) 413 [arXiv:1506.04732].

\bibitem{volume-weighting}D. Page, {\it Space for both no-boundary and tunneling quantum states of the universe, Phys. Rev.} {\bf D 56} (1997) 2065 [gr-qc/9704017];\\
    S.W. Hawking, {\it Volume weighting in the no boundary proposal}, arXiv:0710.2029.

\bibitem{replica}J.B. Hartle and T. Hertog, {\it Replication Regulates Volume Weighting in Quantum Cosmology, Phys. Rev.} {\bf D 80} (2009) 063531 [arXiv:0905.3877].

\bibitem{sad}D. Page, {\it Spacetime Average Density (SAD) Cosmological Measures, JCAP} {\bf 11} (2014) 038 [arXiv:1406.0504].

\bibitem{slih}A. O. Barvinsky and A. Yu. Kamenshchik, {\it Cosmological landscape from nothing: Some like it hot, JCAP} {\bf 09} (2006) 014 [hep-th/0605132]; {\it Thermodynamics via Creation from Nothing: Limiting the Cosmological Constant Landscape, Phys. Rev.} {\bf D 74} (2006) 121502 [hep-th/0611206].

\bibitem{why}A. O. Barvinsky, {\it  	
Why there is something rather than nothing (out of everything)?, Phys. Rev. Lett.} {\bf 99} (2007) 071301 [arXiv:0704.0083].

\bibitem{bigboost}
    A. O. Barvinsky, C. Deffayet and A. Yu. Kamenshchik, {\it Anomaly Driven Cosmology: Big Boost Scenario and AdS/CFT Correspondence, JCAP} {\bf 05} (2008) 020 [arXiv:0801.2063].

\bibitem{DGP/CFT}A. O. Barvinsky, C. Deffayet and A. Yu. Kamenshchik, {\it  	CFT driven cosmology and the DGP/CFT correspondence, JCAP}  {\bf 05} (2010) 034 [arXiv:0912.4604].

\bibitem{CMBA-theorem}A.O. Barvinsky, {\it Thermal power spectrum in the CFT driven cosmology, JCAP} {\bf 10} (2013) 059  [arXiv:1308.4451].

\bibitem{ChibisovMukhanov}
    V. F. Mukhanov and G. V. Chibisov, {\it Quantum Fluctuation and Nonsingular Universe, J. Exp. Theor. Phys.  Lett.}  {\bf 33} (1981) 532;\\
    V. Mukhanov, H. Feldman and R. Brandenberger, {\it  Theory of cosmological perturbations. Part 1. Classical perturbations. Part 2. Quantum theory of perturbations. Part 3. Extensions, Phys. Rept. } {\bf 215} (1992) 203.

\bibitem{whyBFV}A.O.Barvinsky, {\it BRST technique for the cosmological density matrix, JHEP} {\bf 10}  (2013) 051 [arXiv:1308.3270].

\bibitem{universality}L. S. Brown and J. P. Cassidy, {\it Stress Tensors and their Trace Anomalies in Conformally Flat Space-Times, Phys. Rev.}  {\bf D 16} (1977) 1712;\\
    A. Cappelli and A. Coste, {\it On the Stress Tensor of Conformal Field Theories in Higher Dimensions, Nucl. Phys.}  {\bf B 314}  (1989) 707;\\
    I. Antoniadis, P. O. Mazur and E. Mottola, {\it Physical states of the quantum conformal factor, Phys. Rev.}  {\bf D 55} (1997) 4770 [hep-th/9509169];\\
    M. Beccaria and A.A. Tseytlin, {\it Higher spins in $AdS_5$ at one loop: vacuum energy, boundary conformal anomalies and AdS/CFT, JHEP} {\bf 11} (2014) 114 [arXiv:1410.3273].

\bibitem{Veneziano}G. Veneziano, {\em Large-N bounds on, and compositeness limit of, gauge and gravitational interactions}, JHEP 0206, 051 (2002) [arXiv:hep-th/0110129].

\bibitem{cutoff}G. R. Dvali, G. Gabadadze, M. Kolanovic and F. Nitti, {\em Scales of gravity}, Phys. Rev. D 65, 024031 (2002) [arXiv:hep-th/0106058];\\
    G. Dvali and M. Redi, {\em Black Hole Bound on the Number of Species and Quantum Gravity at LHC},  Phys. Rev. D77 ( 2008) 045027 [arXiv:0710.4344];\\
    G. Dvali, {\it Black Holes and Large N Species Solution to the Hierarchy Problem}, Fortsch. Phys. 58:528-536,2010 [arXiv:0706.2050];\\
    G. Dvali and S. N. Solodukhin, {\em Black Hole Entropy and Gravity Cutoff}, CERN-PH-TH-2008-141 [arXiv:0806.3976].

\bibitem{slihinfl}A.O. Barvinsky, A.Yu.Kamenshchik and D.V.Nesterov, {\it Origin of inflation in CFT driven cosmology: $R^2$-gravity and non-minimally coupled inflaton models} [arXiv:1510.06858].

\bibitem{nonminimal}B. L. Spokoiny, {\it  Inflation And Generation Of Perturbations In Broken Symmetric Theory Of Gravity, Phys. Lett.}  {\bf B 147} (1984) 39;\\
    R. Fakir and W. G. Unruh, {\it  Improvement on cosmological chaotic inflation through nonminimal coupling, Phys. Rev.}  {\bf D 41} (1990) 1783;\\
    D. S. Salopek, J. R. Bond and J. M. Bardeen, {\it Designing Density Fluctuation Spectra in Inflation, Phys. Rev.}   {\bf D 40} (1989) 1753;\\
    E. Komatsu and T. Futamase, {\it Complete constraints on a nonminimally coupled chaotic inflationary scenario from the cosmic microwave background, Phys. Rev.}  {\bf D 59} (1999) 064029 [astro-ph/9901127].

\bibitem{Gorbunov}F.L. Bezrukov and D.S. Gorbunov, {\em Distinguishing between R2-inflation and Higgs-inflation}, Phys. Lett. B713 (2014) 365 [arXiv:1111.4397];\\
    D.S. Gorbunov and A.G. Panin, {\it Are $R^2$- and Higgs-inflations really unlikely?, Phys. Lett.} {\bf B 743} (2015) 79 [arXiv:1412.3407].

\bibitem{BezShapSib}F. Bezrukov, A. Magnin, M. Shaposhnikov and S. Sibiryakov, {\it  	
Higgs inflation: consistency and generalisations, JHEP} {\bf 01} (2011) 016  [arXiv:1008.5157].

\bibitem{IR}S. P. Miao and R. P. Woodard, {\it Fine tuning may not be enough} [arXiv:1506.07306].

\bibitem{Vasiliev}M. A. Vasiliev, {\it Consistent equation for interacting gauge fields of all spins in (3+1)-dimensions, Phys. Lett.} {\bf B 243} (1990) 378; {\it Nonlinear equations for symmetric massless higher spin fields in (A)dS(d), Phys. Lett.} {\bf B 567} (2003) 139 [hep-th/0304049].

\bibitem{GKPST}S. Giombi, I. R. Klebanov, S. S. Pufu, B. R. Safdi, and G. Tarnopolsky, {\it AdS Description of Induced Higher-Spin Gauge Theory, JHEP} {\bf 10} (2013) 016 [arXiv:1306.5242].

\bibitem{Tseytlin}A. A. Tseytlin, {\it On partition function and Weyl anomaly of conformal higher spin fields, Nucl. Phys.}  {\bf B 877} (2013) 598  [arXiv:1309.0785].

\bibitem{CHScosmology}A. Barvinsky, {\it CFT driven cosmology and conformal higher spin fields} [arXiv:1511.07625v2].

\bibitem{Komargodski}Z.Komargodski and A.Schwimmer, {\em On Renormalization Group Flows in Four Dimensions}, JHEP 1112 (2011) 099 [arXiv:1107.3987];\\
    Z.Komargodski, {\em The Constraints of Conformal Symmetry on RG Flows}, JHEP 1207 (2012) 069 [arXiv:1112.4538].

\bibitem{CFTCMB}A. Barvinsky and A.Yu.Kamenshchik, work in progress.


\end{thebibliography}
\end{document}